\documentclass[aps,prb,amsmath,amssymb,showpacs]{revtex4}

\usepackage[pdftex]{graphicx}

\begin{document}
  \title{Structures built by steps of evaporated crystal surface Monte Carlo simulations and experimental data for GaN epi layers.}
\author{Magdalena A. Za{\l}uska--Kotur }
\email{zalum@ifpan.edu.pl}
 \affiliation{Institute of Physics, Polish Academy of Sciences,
Al. Lotnik{\'o}w 32/46, 02-668 Warsaw, Poland and Faculty of Mathematics and Natural Sciences,
Card. Stefan Wyszynski University, ul Dewajtis 5, 01-815 Warsaw, Poland}
 \author{Filip Krzy{\.z}ewski} 
\email{fkrzy@ifpan.edu.pl}, 
 \affiliation{Institute of Physics, Polish Academy of Sciences,
Al. Lotnik{\'o}w 32/46, 02-668 Warsaw, Poland} 
\author{Stanis{\l}aw Krukowski} 
\email{stach@unipress.com.pl}
\affiliation{ High Pressure Research Center, Polish Academy of Sciences, ul. Soko\l owska 29/37,01-142 Warsaw, Poland and
Interdisciplinary Centre for Materials Modeling, Warsaw University, Pawi\'nskiego 5a, 02-106 Warsaw, Poland}
 \author{Robert Czernecki and Micha{\l}  Leszczy{\'n}ski} 
\affiliation{ High Pressure Research Center, Polish Academy of Sciences, ul. Soko{\l}owska 29/37,01-142 Warsaw, Poland ,
TopGaN 
Sokolowska 29/37, 01-142 Warsaw, Poland}

\begin{abstract}
We present Monte Carlo simulation data obtained for the annealed surface GaN(0001) and  compare them with the experimental  data. High temperature particle evaporation is a part of  substrate  preparation  processes before epitaxy. The ideal surface ordering expected after such heating is a pattern of parallel, equally distanced steps.  It appears however, that different types of step structures emerge at high temperatures. We show how the creation of characteristic patterns depends on the temperature and the annealing time. The first pattern is created for a very short evaporation time and consists of rough steps. The second pattern built by curly steps is characteristic for longer evaporation times and lower temperatures. The third pattern, in which steps merge together creating bunches of steps happens for the long enough time. At higher temperatures, bunches of steps bend into the wavy-like structures.  
     \end{abstract}
\maketitle

\section{Introduction}
\label{sec:A}

Epitaxial growth of nitride layers, both pure GaN and solid AlGaN and GaInN solutions, by Metal Organic Vapor Phase Epitaxy (MOVPE) or by Molecular Beam Epitaxy (MBE) is the core of nitride based device technology. \cite{nakamura1, skierbiszewski1} Device performance depends on the quality of the synthesized layers, both in terms of crystalographic structure and of their chemical composition. In terms of the growth control, the relative thickness of the deposited layers is relatively small, therefore the results of the growth depend primarily on the structural and chemical state of the surface prior to epitaxy. It is natural that the elaborate surface preparation procedures were developed that are supplemented by  surface characterization techniques.\cite{weyher1, weyher2, weyher3, nowak1}

The epi-ready substrates are mechano-chemically polished to have
atomically smooth surface. \cite{weyher1,pearton,sato} However, when such substrates are introduced
into the epitaxy growth chamber (MOVPE
or MBE ) they contain a variety of molecules
attached when the substrates are exposed into air\cite{monch}. Therefore, prior to the
epitaxial growth, the substrates are annealed to clean the surface\cite{weyher2, weyher3}.
Temperature and time of annealing, as well as ambient atmosphere must be
optimized for every kind of substrate, because the wrong choice of these
parameters may result in a deterioration of the surface.
In this paper, we deal with GaN, the wide-band gap semiconductor used for
fabricating white LEDs, BluRay lasers, and many other devices. The
properties of the GaN crystal differ significantly from other III-V
semiconductors (AlGaIn)(AsP). One of these differences, explored in this
paper, is that GaN crystallizes in the hexagonal wurtzite structure, not in the cubic zinc blende one. In the wurtzite structure, the consecutive atomic steps
of the (0001) surface are different, what has important implication not
only for the growth morphology , but also for the surface changes during
annealing prior the epi-growth. As a result, the difference in the bonding energy at steps  rather than etching anisotropy \cite{garcia,garcia2} leads to the step bunching instability of GaN(0001) surface. 
 In the initial stage of MOVPE growth, the GaN substrate is annealed in the mixture of ammonia, nitrogen and hydrogen, which removes foreign chemical species, and also some portion of gallium nitride. During this stage, the atomistic structure undergoes transformation which could be controlled to some extent by the thermodynamic conditions of the annealing and by duration of the process. 
The main topic of this paper are results of Monte Carlo simulations of the
GaN surface evolution during annealing. These theoretical results are
compared with the experimental data obtained for GaN substrates annealed
in the MOVPE chamber in a flow of ammonia, nitrogen and hydrogen with the
ratio of 2:1:1.

We construct model assuming that all  bonds are saturated by N atoms, hence the process is controlled by desorption and diffusion of Ga atoms only. Lattice gas model of wurtzite structure of GaN is built by Ga atoms, interacting via the layer and orientation dependent forces. This many-body interaction determines the diffusion and desorption rates, which define the kinetics  of the system. Due to the orientation dependent interactions we develop the natural for GaN(0001) surface unequivalency of even and odd step kinetics. We show its consequence for the annealing process at lower and higher temperatures. Surface evolution is presented in this paper along with the Monte Carlo simulations attempting to grasp the nature of the observed changes and the physical forces responsible for the observed scenario. The critical assessment of the obtained results and the conclusions, pertinent to the ammonia based MOVPE homoepitaxial growth of the nitrides, are presented at the end.

\section{The model}
\label{sec:B}
       \begin{figure}
\includegraphics[width=9cm]{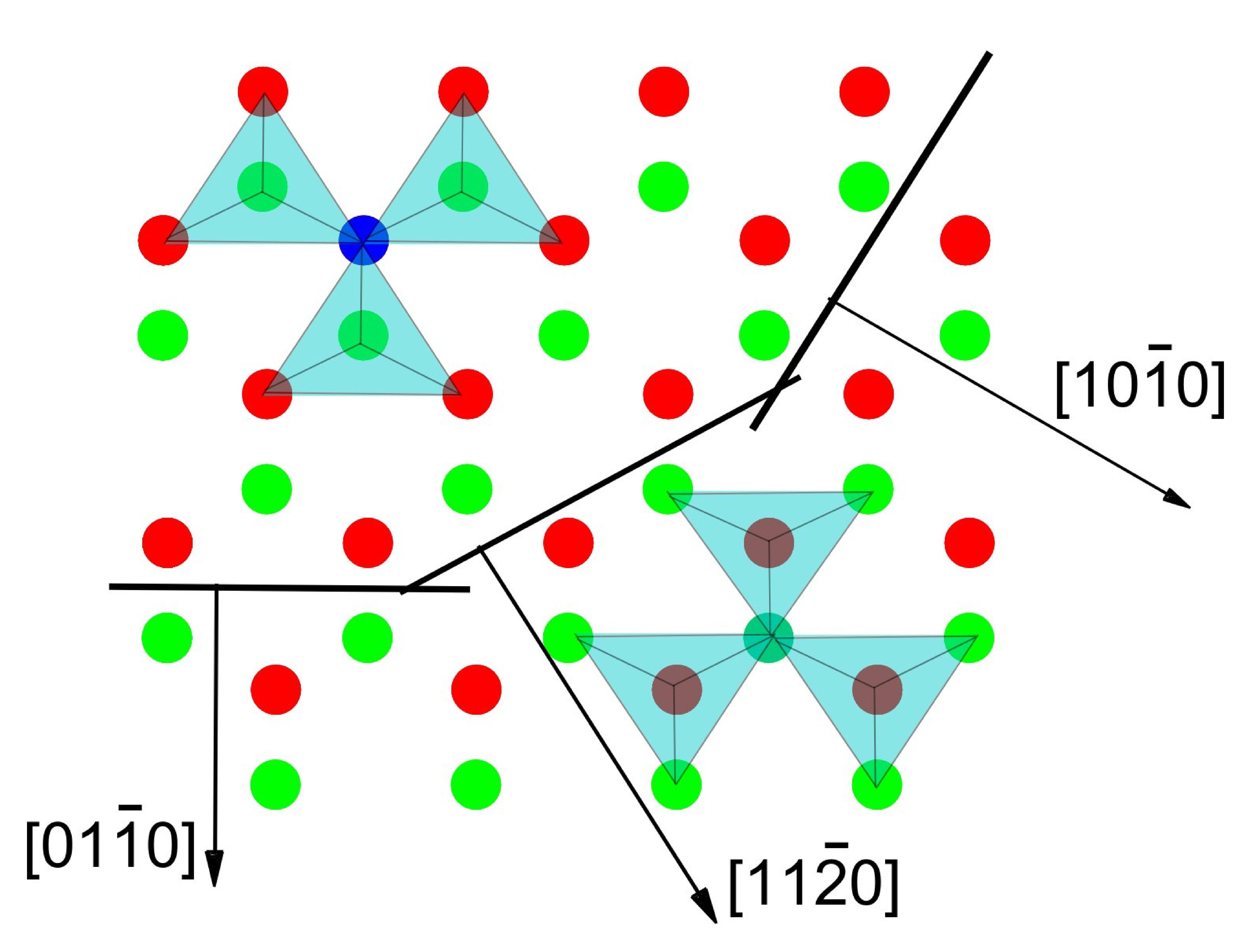}
\includegraphics[width=7cm]{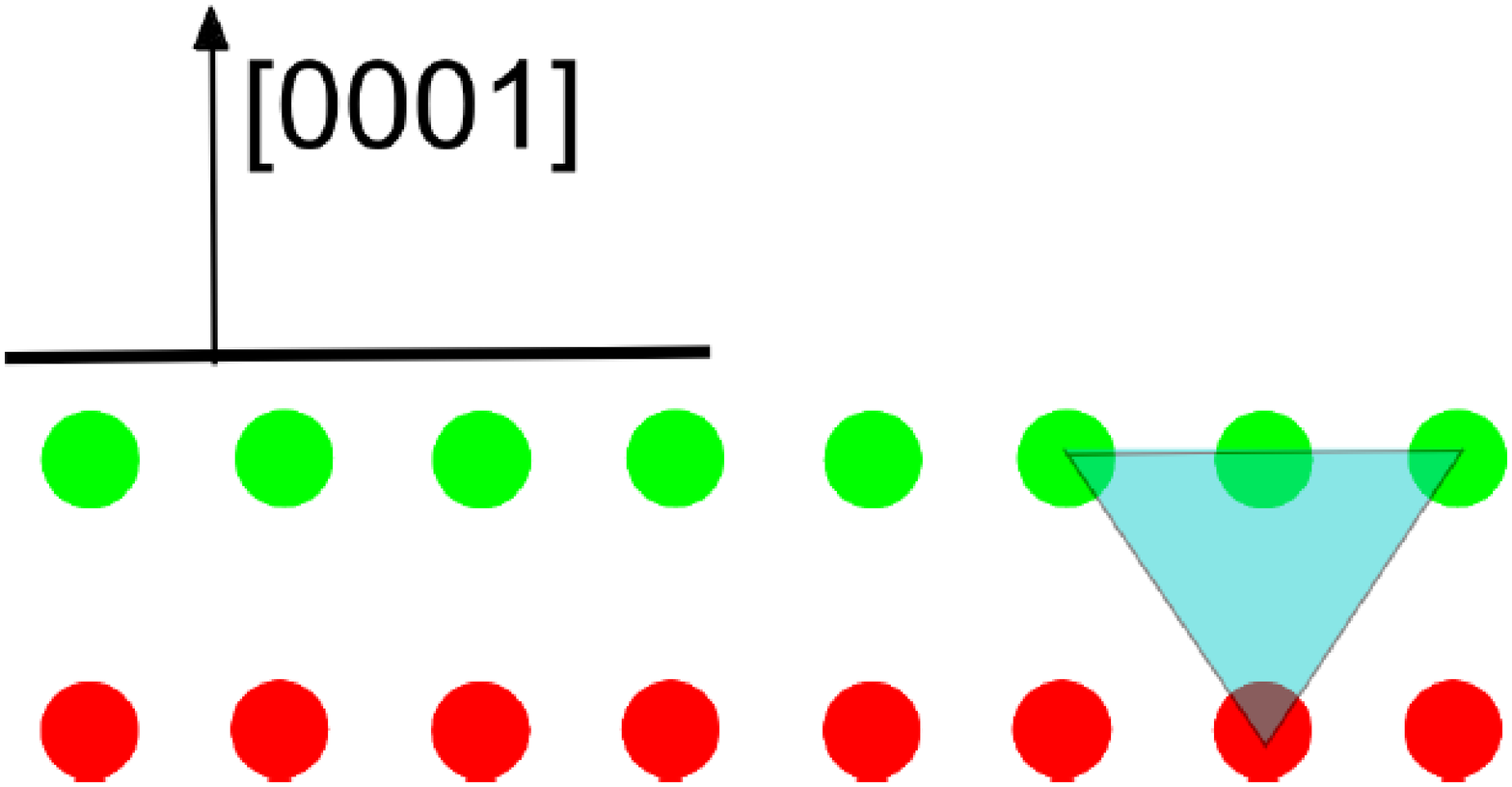}
\caption{\label{model} (color online) Model of GaN crystal. Ga atoms at two  layers are plotted by black (red) and grey (green) circles. Terrace height decreases by one layer at step marked by black line. At upper terrace red particles build  top layer and at lower terrace top layer consists of green particles.  Each tetrahedron contains one N atom inside. Two different bond orientations at following terraces are marked in triangles.  }
\end{figure}
	GaN(0001) surface is built by two consecutive layers, which  differ by the location of Ga sites and by Ga-N  bond orientations. Due to this  factor, two consecutive surface steps have different atomic structure. As illustrated in Fig \ref{model}, three bonds connecting Ga atom with the nearest N atoms within one layer are rotated by 60$^o$ with respect to the bonds in the consecutive Ga layer. As a consequence of this crystallographic difference, step changes its character on coming one terrace down or on the rotation by 60$^o$. 
Such crystallographic structure could result in different energy of bonding to the step. However, when  the structure of the surface of GaN is analyzed more closely, it appears that adatoms at both types of the step have the same number of first neighbors (atoms N) and second neighbors (atoms Ga). Thus, as we have shown in the previous works  \cite{[33],[34]},  the model cannot be limited to the additive two-body interactions only.  The main difference between both steps lies in the relative orientation of the second neighbors of Ga atom, i.e., the closest Ga atoms, namely those from the layer below and those from the step to which our particle attaches. As it has been shown before \cite{[33],[34]}, the simplest way to model such difference is to introduce four-body interaction between Ga atoms. The lattice of the GaN crystal has a wurtzite structure and  therefore, it can be built out of tetrahedrons of Ga atoms centered on N atom. Three atoms of this tetrahedron are visible in Fig \ref{model} around each N atom and the fourth lies below it in the layer underneath. In three-body interaction, each Ga tetrahedron, bonded by single N atom inside has a different energy than the simple sum of two body Ga-Ga bonds, connected by N atom. Note that by assumption of overwhelming presence of nitrogen, we simplified our model, reducing the role of N atoms to the links connecting Ga atoms in the lattice. Accordingly, our system is described by modeling the energy and dynamics of Ga atoms only. We assume that the energy  affecting jump probability of each Ga atom from the lattice site depends on the number of Ga neighbors and on the position of N atom as a tetrahedron center. It is determined in the following way:
\begin{equation}
\label{en_od_trojki}
n_i=\left\{\begin{array}{ll}
1,\quad \textrm{when tetrahedron has all atoms;} \\
\frac{1}{3} r\eta,\quad \textrm{when tetrahedron has empty sites,} \\
\end{array} \right.
\end{equation}
where $\eta$ is a number of occupied neighboring sites, out of those which belong to  a given tetrahedron and $r$ describes the relative  strength of the four-body and the two-body interactions in the system. When $r=1$ two body Ga-Ga interactions sum up to the value characteristic for fully occupied tetrahedron i.e. no additional four-body Ga interactions are present in the system. When $r<1$ three pair bonds to the nearest neighbors of a given particle in tetrahedron do not sum up to a value of one multiparticle bond. In such a case, tetrahedron energy is not a simple sum of two body interactions only and this is the case studied. The value $r=0.4$ is used all along the work. 

At GaN(0001) surface, each Ga surface atom  belongs potentially to four tetrahedrons, the three in the  present layer and the one above. Its total energy we calculate as
\begin{equation}
\label{en_czastki}
E(J)=J\sum_{i=1}^{4}n_i.
\end{equation}
Parameter $J$ scales energy of bonds, and the sum runs over four tetrahedrons, that surround every atom. 
 
We study GaN crystal surface dynamics by controlling kinetics of Ga atoms. At given temperature particles can jump between adsorption sites. Probability of a jump from the initial to the final site, in the  is given by parameter
\begin{equation}
\label{p_d}
D=D_0 e^{-\beta \Delta E},
\end{equation}
where $D_0=1$ is the diffusion timescale, $\beta=1/k_BT$ is the parameter of the inverse temperature, $\Delta E$ depends on the initial $E_i(J)$ and the final $E_f(J)$ bonding energy of the jumping  atom,
\begin{equation}
\label{eq:E}
\Delta E=\left\{\begin{array}{ll}
0,\quad \textrm{if $E_i(J)<E_f(J)$;} \\
E_i(J)-E_f(J),\quad \textrm{otherwise.} \\
\end{array} \right.
\end{equation}
Particles, which are adsorbed at the step can detach with probability (\ref{p_d}), which depends on the energy of particle bonding to the step via (\ref{eq:E}). When particle is free at the terrace it diffuses over  the surface until it escapes out of it. Desorption probability is given by
\begin{equation}
p_d=a_d e^{-\beta \nu}
\end{equation}
with timescale $a_d$, which is set one and the desorption potential $\nu$. The desorption rate is controlled via temperature $\beta$ and potential $\nu$.
Readsorption at the step is determined by the rate with which particle once detached from the step jumps back into it. This rate is additionally modified by Schwoebel barrier \cite{Schwoebel} that sets up different probability when atoms diffuse to the step from upper and lower terrace. The height of the barrier is described by the parameter $B$. Probability of the jump over the barrier is as follows:
\begin{equation}
\label{p_d^B}
D_B=e^{-\beta B} D.
\end{equation}
thus, the height of the barrier $B$ modifies jump rate given by equation (\ref{p_d}). As typically assumed, we impose Schwoebel barrier for the jumps from and to the upper terrace. We use the same barrier height  $B$ for both types of steps. 

Crystal surface microstate is modeled by setting two uppermost layers of atoms. In all simulations, the surface  is misoriented along one direction. When new step appears, on coming up, the upper layer is converted into the lower one, and a new layer is built on top of the terrace. In such a way, a continuity of particle-particle interaction at the step is guaranteed. 
Every second layer of Ga atoms has different bond orientation. Due to different positions of N atoms, orientation of tetrahedral bonds is rotated by 60$^o$ in alternate layers. Such geometry causes step flow anisotropy characteristic for GaN surface. 

We start our simulations with an  even  number of equally spaced by $d$ lattice constants, straight steps  on the surface. Heights of the neighboring steps differ by one Ga atomic layer. Ga atoms detach from the step, diffuse along the terrace and desorb from the surface. Crystal height decays layer by layer. Periodic boundary conditions are applied in the lateral direction and in the direction in which the crystal grows they are corrected by constant height difference between both ends of the system.
\begin{figure}
\includegraphics[width=9cm]{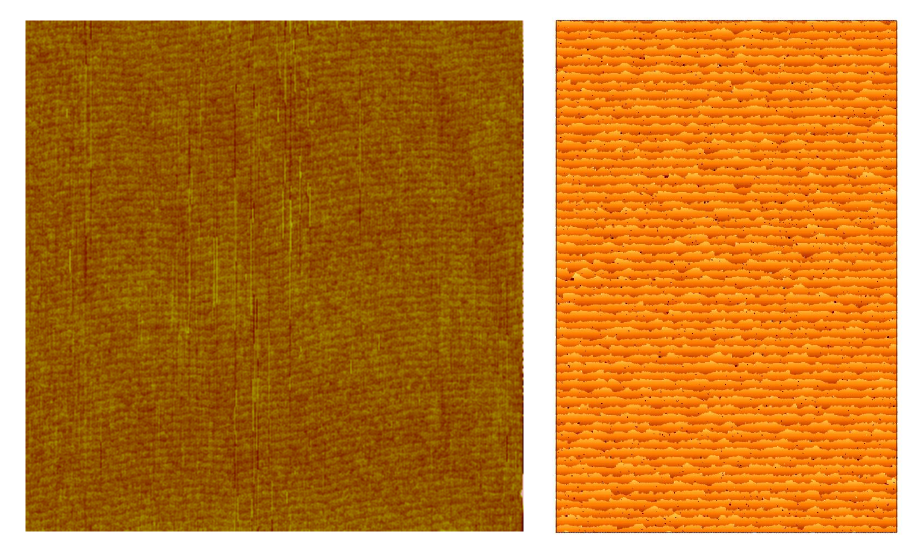}
\caption{(color online) At the left hand side substrate prepared during mechanochemical cutting of the crystal is shown. Cutting angle is $1^o$, and step direction is $[11\bar{2}0]$ . Steps are straight,  parallel but rough. The same effect we can see in the simulated system in the initial stage of the evaporation process as shown at the right side of the plot. Size of the simulated system was $400a \times 600a$, $k_BT=0.2J$ and initial terrace width $d=10a$, $r=0.4$ , $B=0.5$ and $\nu=1.6J$. 
} 
\end{figure}

\begin{figure}
\includegraphics[width=9cm,angle=0]{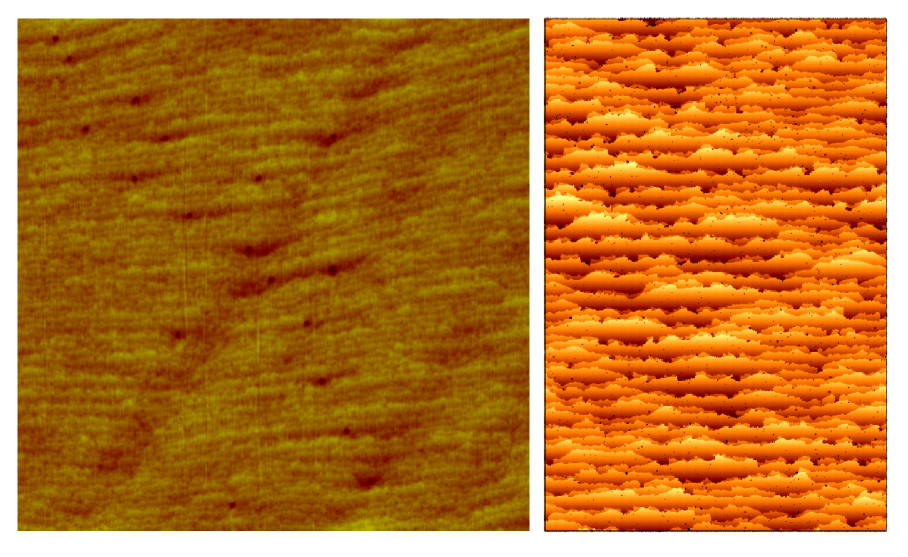}
\caption{\label{curly} (color online) At left hand side GaN (0001)surface of miscut $1^o$, where steps are oriented along direction $[11\bar{2}0]$ after heating at $T=900^oC$ shows characteristic curly structure. Steps are bent in alternate way giving expression that terraces broaden.  
Step pattern of the simulated system  evaporated  at $k_B T=0.2J$ with steps  $[11\bar{2}0]$  and initially separated by $d=10 a$ distance is shown in the right panel.  Simulation was carried out for system of size $400 \times 600$ lattice constants, $r=0.4$ , $B=0.5$ and $\nu=1.6J$. Step evolution, after $2.5^.10^6$ MC steps, is presented at the right side, that is equivalent to several minutes of the experimental time.}
\end{figure}

\begin{figure}
\includegraphics[width=9cm,angle=0]{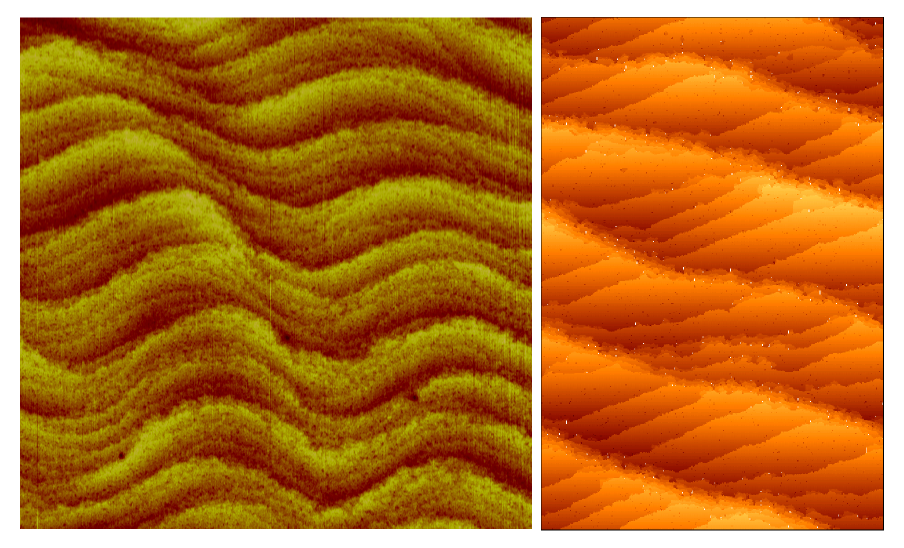}
\caption{\label{wavy} (color online) At left hand side GaN (0001)surface of miscut $1^o$ and steps oriented along $[11\bar{2}0]$ after heating at $T=1050^oC$ shows wavy structure.  
Step pattern which evolves towards meandered step train structure is shown in the system evaporated at $k_B T=0.23J$ with steps  initially oriented along $[11\bar{2}0]$  and separated by $10 a$.  Simulation pattern obtained for system of size $400 \times 600$ lattice constants, $r=0.4$ , $B=0.5$ and $\nu=1.6J$ after $10^7$ MC steps. At this temperature, it is equivalent to several minutes of evaporation.}
\end{figure}

\begin{figure}
\includegraphics[width=11cm]{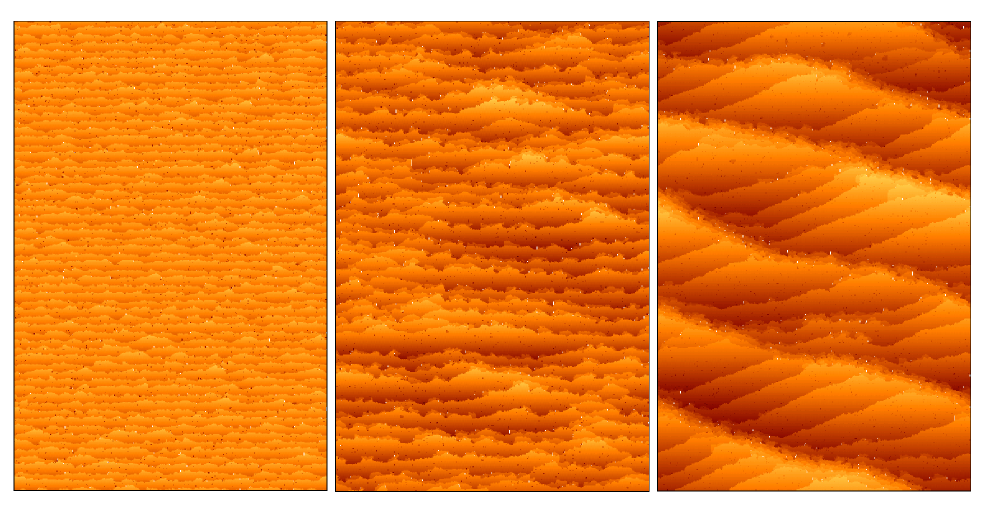}
\caption{Step pattern build in the evolution of surface evaporated at  the temperature $k_BT =0.23J$. Patterns obtained after $10^4$, $2.5 10^6$ and $10^7$ MC steps, are shown.   Other parameters are the same as in Fig \ref{wavy}
} 
\end{figure}

\section{ Step pattern evolution in the etching  process. }
\label{sec:B2}

Surface is heated, and kept at the high temperature. In the simulation process step patterns transform from one form to the other depending on the simulation time.	We start from the regular pattern of equally spaced steps. At the beginning steps are straight and rough. In left panel of Figure 2. we show the surface  as it is prepared by mechanochemical polishing before evaporation. It can be compared with the initial phase of simulations in right panel of Figure 2, when primary smooth steps become rough, but they are still straight. Behavior of steps both in the experimental and simulated situations  have the same origin. During mechanochemical procedure particles, which are weakly bonded detach from the steps. Temperature is too low to activate process of the diffusion along steps. The  same situation happens in the beginning of the heating process of straight steps. Weakly bonded particles detach, and time is too short for the diffusion to play role in the ordering process. The only step smoothening, which happens takes place only for very short distances. As a result both patterns look similarly.   

 Next stage of the step evolution is heating. Starting from the discussed surface ordering,  different surface patterns are created, depending on the temperature and the evaporation time. The structure, which is characteristic  for lower temperatures consists of steps characteristically bent creating specific curly pattern. This pattern from above looks like the system of parallel lines, similar to the initial one, but more sharp and locally  wider. Illustration for such ordering can be seen in Figure 3, in the left panel for the experimental system heated at 900$^oC$, as well as for the simulated surface in the right panel. In simulations we used parameter $k_BT=0.2J$ what corresponds to the  experimental 900$^oC$, assuming that the mean activation energy for particle detach from the step is equal to $0.7 eV$. Desorption rates from the surface were defined by the potential $\nu$. The value of parameter, which gave the results most close to the experimental ones was $1.6J$, what translates to the barrier for desorption of the single particle from the surface being equal to $1.12eV$. Both these values are quite close to the experiment. As we wanted to simulate large systems within the real experimental period of time, our simulated crystal is prepared with terraces of width $10a$ where $a$ is lattice constant. This is smaller value than $60a$, corresponding to the experimental miscut of $1^o$.
When we simulate system of higher terraces width, our results are similar, but then the sizes of the systems we can investigate are much smaller. Hence, we show results for a higher miscut than in the experiment, but of system sizes and simulation times comparable to the experimental ones. 

 The same system has been evaporated at higher temperature. In Figure 4, we can  see GaN surface of the same orientation and cut as in Figure 3, but after heating at 1050$^oC$. Time of heating was the same as in the previous discussed case, what means several minutes, but resulting surface pattern is completely different. Instead of the structure of parallel alternatively bent lines characteristic for lower temperatures, at higher temperatures we can see step bunches which form large waves.   The same pattern was created in simulations at $k_BT=0.23J$, which corresponds to $T=1050^oC$ and simulation times comparable to that in the experiment. In the right panel of Figure 4 we see waves of step bunches. 

    We run many simulations at different temperatures. The conclusion is that the typical development of evaporated systems occurs as follows. For short evaporation time only part of the surface layer is removed and steps move onto the small distances only and become rough as we see in Figure 2. On moving further, steps bend on longer distance forming characteristic curly structures like this in Figure 3. When the system is  evaporated for longer time steps have tendency to group into step trains. It can be seen in the  Figure 4. Figure 5 shows the described above system evolution observed at higher temperature $k_BT=0.23$. At such temperature evolution process is faster. We find step bunching in the systems of various misorientations, step directions and temperatures. All these parameters determine how many MC steps are needed to obtain such structures. Moreover, when step trains are created easily, as it is observed at higher temperatures, then after creation, such steps start to bend in the higher scale, developing wavy-like structures. We presented such structures in the left panel of Figure 4, emerging in the system evaporated at 1050$^o$C and comparable  structures obtained in the MC simulated system are shown in the right panel. It has to be noted that some of the structures, described above, have different character than structures built during crystal growth \cite{[33],[34]}. The main discrepancy lies in the tendency  of forming bunches. Simulated surfaces of growing crystals rather build single step patterns, than create step bunches. During evaporation steps group much more easily. More detailed comparison of both processes would be instrumental in understanding surface dynamics more precisely and such investigations are presently carried out.

\section{Conclusions}

We studied step patterns created in the process of temperature etching of the GaN surface. Comparison of the simulated systems with ordering of real, experimental  GaN(0001) surfaces heated at different temperatures shows similar behavior. The main types of experimentally observed surface ordering were   reproduced by lattice model of GaN crystal with many body interactions. The tendency of the formation of concrete surface patterns depends strongly on the temperature. When heating temperature is too high, steps tend to group in bunches and form wavy patterns. There seems to exist only small window of temperatures, where crystal, on heating has parallel, equally distanced step, bent a little in a characteristic curly structure. Such structure is the best initial state for growing regular crystal layers.  On the other side we observed also that evaporation process has not the same dynamics as growth with reversed flux. Even when we compare systems grown and etched  with the same particle flux in and out, and at the same temperature, still the resulting structures are different. The cases we studied up to now prove that steps at GaN(0001) surface bunch much more easily during etching than during growing process. The structures, they eventually form, look differently.

\label{sec:D}

\section{Acknowledgement}
 Research supported by the European Union within European Regional Development Fund, through grant Innovative Economy (POIG.01.01.02-00-008/08)

\end{document}